\documentclass[preprint,12pt]{elsarticle}

\usepackage{epsfig}

\usepackage{graphicx}

\usepackage{url}

\usepackage{amssymb}

\newcommand{\C}{\mbox{$\mathbb C$}}

\journal{Computer Physics Communications}

\begin{document}

\begin{frontmatter}



\title{A derivative-free optimization method to compute scalar perturbation of AdS black holes}


\author{R. S. Lima\corref{cor1}}
\ead{rodlima@unifei.edu.br}
\cortext[cor1]{corresponding author: Tel. +55 35 3629 1435}
\address{Institute of Mathematics and Computation,
           Federal University of Itajub\'a,
           Itajub\'a, Brazil}
 \author{A. B. Pavan\corref{cor2}}
 \address{Institute of Physics and Chemistry,
           Federal University of Itajub\'a,
           Itajub\'a, Brazil}

\begin{abstract}
In this work we describe an interesting application of a simple derivative-free optimization method to extract the quasinormal modes (QNM's) of a massive scalar field propagating in a 4-dimensional Schwarzschild anti-de Sitter black hole (Sch-AdS$_4$). In this approach, the problem to find the QNM's is reduced to minimize a real valued function of two variables and does not require any information about derivatives. In fact, our strategy requires only evaluations of the objective function to search global minimizers of the optimization problem. Firstly, numerical experiments were performed to find the QNM's of a massless scalar field propagating in intermediate and large Sch-AdS$_4$ black holes. The performance of this optimization algorithm was compared with other numerical methods used in previous works. Our results showed to be in good agreement with those obtained previously. Finally, the massive scalar field case and its QNM's were also obtained and discussed.
\end{abstract}

\begin{keyword}
black holes \sep scalar perturbation \sep nonlinear optimization \sep derivative-free method \\[8pt]

\PACS 04.50.Gh \sep 02.60-x

\end{keyword}

\end{frontmatter}


%%%%%%%%%%%%%%%%%%%%%%%%%%%%%%%%%%%%%%%%%%%%
\section{Introduction}\label{intro}

Black holes are probably the most exotic gravitational objects of the General Relativity Theory. From the theoretical physics point of view, a black hole is one type of solution of Einstein equations whose the main characteristic is the existence of an one-way hypersurface that once something crosses it, can not return, even the light, thus defining the event horizon. On the other hand, from the astrophysical point of view black holes are one possible final stage of collapsing stars. Actually the black hole modeling given by General Relativity is the best mathematical description of these objects.

The first black hole solution obtained was a static, spherically symmetric spacetime \cite{schwarzschild} whose the event horizon is located at $r=2\mathcal{M}$. It was named Schwarzschild black hole. After this solution, generalizations also was obtained such that a static spherically symmetric electrically charged black hole called Reissner-N\"ordstrom black hole \cite{ReissnerNordstrom}, the Schwarzschild-de Sitter black hole and Schwarzschild anti-de Sitter both static spherically symmetric black hole with cosmological constant $\Lambda>0$ and $\Lambda<0$, respectively. Many other black hole solutions were found in the context of alternative theories of gravitation and in higher \cite{horowitzbook} and lower dimensions \cite{BTZ}. Other important black hole solutions are the Kerr black hole a stationary rotating black hole \cite{Kerr} and a stationary rotating electrically charged named Kerr-Newman black hole \cite{kerrnewman}.

As important as finding a black hole solution is to analyze its stability when submitted to perturbations. Gravitational perturbations in black hole solutions was firstly studied by Regge and Wheeler that analyzed a small perturbations $h_{\mu\nu}$ in the metric $g_{\mu\nu}$ of Schwarzschild black hole and found its stability \cite{ReggeWheeler}. Electromagnetic, fermionic and scalar perturbations also can be performed in order to test the stability of the black hole in an indirect way \cite{chandrasekhar,Ruffini}. In these cases, the spacetime is keeping fixed while the field is permitted evolve. If the temporal evolution of the field is damped in time this indicates stability of the black hole under that perturbation. This damping can be investigated by means of the quasinormal modes (QNM) of the field, i.e., complex frequencies of oscillation of the field propagating in a black hole spacetime. For additional information about QNM's see the following reviews \cite{Nollert,Kokkotas}. Our previous experience on black hole perturbations in many different contexts can be seen in \cite{alan1,alan2,alan3,alan4}.

The study of the quasinormal modes is very important to understand the dynamic of black holes because they are strongly related to their basic properties such as $\mathcal{M}$ mass, $Q$ electrical charge and $J$ angular momentum. However, the task of computing these QNM's is very hard. In general, a non-linear partial differential equation involving the physical characteristics of the black hole and matter fields is solved by a numerical method since rarely exact solution can be obtained. In two important works about this topic \cite{horowitz,konoplya}, the authors extract the QNM's of a massless scalar field evolving in Schwarzschild-AdS$_d$ black hole using built-in functions of the software \emph{Mathematica} (\url{http://www.wolfram.com/mathematica}). However, there are few discussions about the performance of the software and the numerical difficulties in their works.

In this work we deal with QNM's of a massive scalar field evolving in Schwarzschild-AdS$_4$ black hole applying a derivative-free optimization method. The method is based on the well-known Luus-Jaakola algorithm \cite{luus1} \-and requires only evaluations of the objective function to search global minimizers of the problem. We carry out some numerical experiments using this approach and compare the results with those obtained in the literature. Our paper is organized as follows: in the Section \ref{sec2} a briefly description of the mathematical tools to model the evolution of a scalar field propagating in the black hole spacetime and the classical procedure to compute its quasinormal modes are presented. In the Section \ref{sec3} the numerical optimization method is shown, explaining how it can be implemented to solve the problem. The results of the numerical experiments performed are discussed. Finally, in the Section \ref{sec4} our conclusions and some ideas to a future work are presented.

%%%%%%%%%%%%%%%%%%%%%%%%%%%%%%%%%%%%%%%%%%%%
\section{Schwarzschild-AdS$_d$ black hole: mathematical background} \label{sec2}

Recently, the interests on black hole perturbation have been renewed because of the advent of AdS/CFT correspondence \cite{adscft}. This correspondence connects holographically a $(d+1)$-dimensional AdS spacetime to a $d$-dimensional conformal quantum field theory living on the boundary of  that spacetime. In this context, perturbations in asymptotically AdS black hole play an important role whereas they are related to phase transitions and linear response of the dual system on the boundary. One well succeed example of the AdS/CFT is the holographic superconductor. In this case, the correspondence establishes a relation between a massive charged scalar field coupled to a Maxwell field propagating in Schwarzschild-AdS$_5$ and quantum description of superconductor. For a good review see \cite{3hs}.

We are interested in explore the evolution of massive scalar field evolving in $d$-dimensional Schwarzschild-AdS black hole whose metric is
\begin{equation}
\label{metricd}
ds^2 = -A(r)dt^2+\frac{1}{A(r)}dr^2+r^2\ d\Omega_{d-2}^2,
\end{equation}
where
\begin{equation}
\label{fr}
A(r) = 1 - \left(\frac{r_{0}}{r}\right)^{d-3} + \frac{r^2}{R^2}.
\end{equation}
This black hole solution is characterized by the anti-de Sitter radius $R$ related to a cosmological constant $\Lambda<0$ on
\begin{equation}
\label{adsr}
R=\sqrt{\frac{(d-1)(d-2)}{-\Lambda}}
\end{equation}
and the black hole mass $\mathcal{M}$ related to $r_0$ on
\begin{equation}
\label{mass}
\mathcal{M}=\frac{(d-2)r_0^{d-3}}{16\pi G_d} \left\{\frac{2\pi^{(d-1)/2}}{\Gamma\left[(d-1)/2\right]}\right\},
\end{equation}
where $G_d$ is the $d$-dimensional Newton's constant.

As one knows, the evolution of a massive scalar field $\Psi$ is driving by Klein-Gordon equation
\begin{equation}
\Box \Psi - m^2 \Psi = 0,
\end{equation}
where $m$ is the mass of the scalar field. Then, if the D'Alembertian $\Box$ is expanded, it leads to the following equation of motion for $\Psi$
\begin{equation}
\label{KGequation}
\frac{1}{\sqrt{-g}}\ \partial_\mu \left[\sqrt{-g}g^{\mu\nu}\partial_\nu \Psi \right]-m^2\Psi = 0,
\end{equation}
where $g$ is the determinant of the metric.

A classical and well-known procedure to calculate the QNM's for scalar perturbation in asymptotically AdS black holes was proposed by Horowitz and Hubeny in \cite{horowitz}. Here we will extend their procedure to massive scalar field case. Rewriting the metric, Eq. (\ref{metricd}), in ingoing Eddington-Finkelstein coordinate $v=t+r_*$ , where $dr_* =\frac{1}{A}dr$ is a new radial coordinate, leads to
\begin{equation}
\label{metricEFd}
ds^2 = -A(r)dv^2+2drdv+r^2\ d\Omega_{d-2}^2.
\end{equation}
In this new system of coordinates, the Eq.(\ref{KGequation}), can be reduced to an ordinary differential equation by the following separation of variables:
\begin{equation}
\label{separation}
\Psi(t,r,\mbox{angles})=\frac{Z(r)}{r^{(d-2)/2}}\ e^{-i\omega v}\ Y(\mbox{angles}),
\end{equation}
where the higher-dimensional angular function $Y(\mbox{angles})$ denotes the spherical harmonics on $S^{d-2}$. If we set $(R=1)$ the ordinary differential equation resultant will be
\begin{equation}
\label{eqdif1}
A(r) \frac{d^2Z(r)}{dr^2}+ \left[~A'(r)-2i \omega ~\right]\frac{dZ(r)}{dr} -V(r) Z(r)=0,
\end{equation}
where the effective potential $V(r)$ is
\begin{equation}
\label{pot}
V(r)=\frac{d(d-2)}{4} + \frac{(d-2)(d-4) + 4c}{4r^2} + \frac{(d-2)^2 r_0^{d-3}}{4r^{d-1}}+m^2,
\end{equation}
and $c=\ell_*(\ell_* +d-3)$ is the eigenvalue of the Laplacian on $S^{d-2}$. In order to compute the QNM's we need a solution in a power series about the horizon and impose the
boundary condition such that the solution vanish at infinity since the effective potential is divergent when $r\to\infty$. To achieve this purpose, we perform another change of coordinates $x=1/r$ in Eq. (\ref{eqdif1}). Thus, it can be rewrite as follows
%,
\begin{equation}
\label{eqHHx}
s(x)\ \frac{d^2Z(x)}{dx^2}+\frac{t(x)}{(x-x_{+})}\ \frac{dZ(x)}{dx}+\frac{u(x)}{(x-x_{+})^2}\ Z(x)=0,
\end{equation}
where the functions $s(x)$, $u(x)$ and $t(x)$ are
\begin{eqnarray}
\label{fHHx}
s(x)&=& \frac{r_0^{d-3}x^{d+1}-x^4-x^2}{x-x_+} ,\\[10pt]
t(x)&=& (d-1) r_0^{d-3} x^d - 2x^3 - 2x^2 i \omega, \\[10pt]
u(x)&=& (x-x_{+}) V (x),
\end{eqnarray}
and the parameters $r_+$ and $r_0$ will be
\begin{equation}
\label{horizonx}
x_+ =  \frac{1}{r^+},\qquad r_{0}^{d-3}=\frac{x_+^{2}+1}{x_+^{d-1}}.
\end{equation}
Since $s(x)$, $t(x)$ and $u(x)$ are polynomials of degree $d$, it is useful expand them about the horizon $x_+$ like
\begin{equation}
s(x) = \sum_{n = 0}^{d} s_n (x - x_+)^n,
\end{equation}
and similarly for $t(x)$ and $u(x)$. Then, a solution to the Eq. (\ref{eqHHx}) in power series is obtained by expanding the function $Z(x)$ near the horizon $x_+$:
\begin{equation}
Z(x)=\sum_{n=0}^{\infty}a_{n}(\omega)(x-x_{+})^{n}. \label{solution}
\end{equation}
Substituting this solution in the Eq. (\ref{eqHHx}) and doing some algebraic manipulation we found the following recurrence relation for $a_n$
\begin{equation}
a_{n}(\omega) = - \frac{1}{P_n} \sum_{k = 0}^{n-1} (~ k(k-1) s_{n-k} + kt_{n-k} + u_{n-k} ~) a_k
\end{equation}
with
\begin{equation}
P_{n}=n(n-1)s_{0}+nt_{0}.
\end{equation}

Since we are interested in normalizable modes we have to select only solutions which satisfy the Dirichlet boundary conditions $Z(x) \to 0$ as $x \rightarrow 0$. These conditions are satisfied only for specific values of $\omega$. Indeed, they transform the calculation of $\omega$ in a problem to find roots of the equation $Z(0) = 0$ in the complex $\omega$ plane. In general, the previous authors solve the problem using the software Mathematica.
They truncate the series $Z(0)$ and solve the equation $Z(0) = 0$ using the routine \emph{FindRoot}. Other approach consists in minimize the function $|Z(0)|$ via routine \emph{FindMinimum}.
Because these routines are, in some sense, black boxes, we do not have complete control of their parameters. Also, the task of determine expressions to the first derivatives of the function $Z(0)$ is very difficult. To find the values of $\omega$, we propose to solve a box constrained optimization problem. In the next section, we describe a derivative-free method to circumvent the numerical difficulty in to calculate the QNM's.

%%%%%%%%%%%%%%%%%%%%%%%%%%%%%%%%%%%%%%%%%%%%
\section{Numerical optimization method}\label{sec3}

In order to obtain the QNM's we need to truncate the series in the Eq.(\ref{solution}) at $x = 0$ to obtain the complex function $Z_N: \C \longrightarrow \C$ given as
\begin{equation}
Z_N (\omega) = \sum_{n = 0}^N~a_n(\omega)(-x_{+})^n, \label{fzn}
\end{equation}
where $x_{+}$ is a real constant and $N$ is a large but fixed integer. We are concerned in find a complex number $\omega \in \C$ in such a way that $Z_N(\omega) = 0$. To attain this goal, we propose to solve the following optimization problem with box constraints
\begin{equation}
\begin{array}{ll}
\mbox{minimize} & f(w) =  |Z_N(w)|\\[5pt]
\mbox{subject to} & \ell \leq \omega \leq u, \label{prob_otim}
\end{array}
\end{equation}
where $\ell$ and $u$ are, respectively, lower and upper bounds to $\omega$ in the complex plane.

The real valued function $f(\omega)$ in (\ref{prob_otim}) attains only nonnegative values because it is defined as the norm of the complex number $Z_N(w)$. Then, if we determine $\omega^* \in \C$ such that $f(\omega^*) = 0$ with $\ell \leq \omega^* \leq u$, the number $\omega^*$ will be a global minimizer of the optimization problem (\ref{prob_otim}). It is easy to verify that if $\omega^*$ is a global minimizer of $f(\omega)$, then $\omega^*$ is also a root to the nonlinear equation $Z_N(\omega) = 0$. In particular, we are interested in find $w^*$ close to the origin with positive imaginary part.

We employ a derivative-free optimization method based on the \emph{Luus-Jaakola algorithm} \cite{luus1} to solve the problem (\ref{prob_otim}). This approach is an attractive for global optimization problems due to three main reasons: (a) the capacity of escape from local minimizers and find solutions in the proximity of global ones; (b) the easiness of implementation in any computational language and (c) only evaluations of function are needed to search candidates to solutions. On the other hand, the main drawbacks are: (a)  the algorithm does not guarantee global optimality, nevertheless some works proved their ability to reach the best known solutions and (b) it may require a large number of function evaluations. Successful applications of this numerical method can be found in \cite{lobato,luus2,luus3}. Rich theory about derivative-free optimization problems and numerical methods of this kind can be found in \cite{vicente,lucas,nocedal}.

It is very simple to describe the search procedure of the method. We begin selecting any $\omega_0$ in the box $[\ell, u]$ at random. Then, we search for a point $w_1$ in the neighbourhood of $w_0$ that attains the lowest value of function $f(w)$. For this, a region $B_0$ centered in $\omega_0$ is constructed and a set of points $\omega_j$ is generated inside $B_0$. Typically, this region can be a ball with radius $\rho_0 > 0$. Let $\omega_1 \in B_0$ such that $f(\omega_1) \leq f(\omega)$ for all $\omega \in B_0$. If $f(\omega_1)$ is lower than a threshold $\epsilon > 0$, we accept $\omega_1$ as a solution to the problem (\ref{prob_otim}). Otherwise, we create a new region $B_1$ centered in $\omega_1$ with radius $0 < \rho_1 < \rho_0$ and repeat the search inside $B_1$. The main steps of this approach are summarized below:
\newpage
\begin{itemize}
\item[] Given a large integer $M$ and a tolerance $\epsilon > 0$:
\item[] {\bf Step 1:} Select $\omega_0$ in $[\ell, u]$, compute $f(\omega_0)$ and do $k \leftarrow 0$.
\item[] {\bf Step 2:} While $k \leq M$: construct a ball $B_0$ centered in $\omega_0$ with radius $\rho_0$.
\item[] {\bf Step 3:} Generate a set of $m$ points inside $B_0 \cap [\ell, u]$.
\item[] {\bf Step 4:} For each point $\omega_j \in B_0 \cap [\ell, u]$: if $f(\omega_j) < f(\omega_0)$, then $\omega_0 \leftarrow \omega_j$.
\item[] {\bf Step 5:} If $f(\omega_0) < \epsilon$, then print $\omega_0$ and stop the algorithm.
\item[] {\bf Step 6:} Otherwise: reduce the radius $\rho_0$, do $k \leftarrow k + 1$ and go to {\bf Step 2}.
\end{itemize}

In the algorithm above, the parameters $M$ and $m$ are, respectively, the maximum number of outer and inner iterations. It means that the total number of evaluations of function is less or equal than $Mm$. The Figure \ref{fig1} below shows two iterations of the method to search the solution $\omega^*$ (green point).
\begin{figure}[htbp]
\centering
\setlength{\unitlength}{1cm}
\begin{picture}(6,6)(-3,-3)
\put(0,0){\makebox(0,0)[cc]{\includegraphics[height = 6cm]{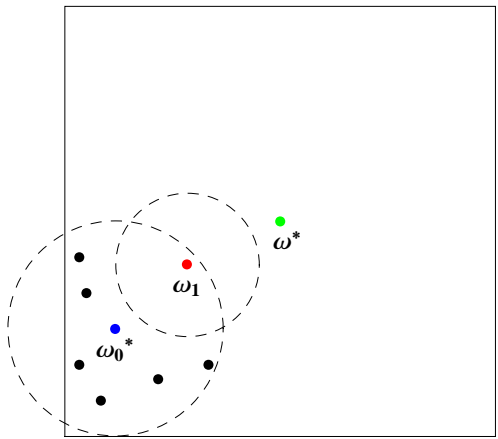}}}
\end{picture}
\caption{Two iterations of the optimization method.}
\label{fig1}
\end{figure}

Given a ball $B_0$ centered in $\omega_0$ (blue point) with radius $\rho_0 > 0$, the algorithm generates a set of points inside $B_0 \cap [\ell, u]$ and evaluates the function $f(\omega)$ in all of these points until find $\omega_1 \in B_0 \cap [\ell, u]$ (red point) such that $f(\omega_1) < f(\omega)$ for all $\omega \in B_0\cap [\ell, u]$. Then, a new ball centered in $\omega_1$ with radius $\rho_1 < \rho_0$ is created and the search procedure is repeated.
%

%
%

%%%%%%%%%%%%%%%%%%%%%%%%%%%%%%%%%%%%%%%%%%%%
\section{Numerical experiments}\label{sec4}

To illustrate the algorithm proposed, we focus our attention in to investigate the QNM's of a massive scalar field evolving in a Schwarzschild-AdS$_4$ black hole. In this case, the metric given by the Eq. (\ref{metricEFd}) becomes
\begin{equation}
\label{metric4d}
ds^2 = -A(r)dv^2+2 dvdr+r^2\ d\theta^2+r^2\sin\theta^2 d\phi^2,
\end{equation}
where
\begin{equation}
\label{fr4}
A(r) = 1 - \frac{2\mathcal{M}}{r} + r^2
\end{equation}
and we have set $R=1$ without any loss of generality. The event horizon is located at $r_+$ such that $A(r_+)=0$. This black hole can be classified in three different types according to its size in relation of $R$, namely, small if $(r_+\ll 1)$, intermediate if $(r_+ \sim 1)$ and large if $(r_+\gg 1)$.
\begin{figure}[htbp]
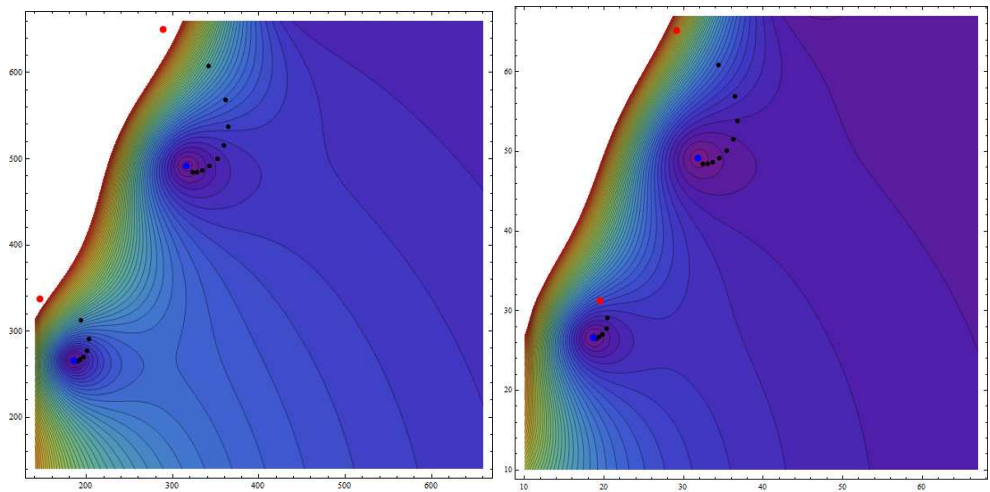

\begin{center}
\epsfig{file =CurvaNivel_num_r100_color.eps, angle=0, width=0.47 \linewidth}
\epsfig{file =CurvaNivel_num_r10_color.eps, angle=0, width=0.47 \linewidth}
\caption{Contour levels and evolution of $f(\omega)$ minima as function of $N$ for $r_+ =100$ (left) and $r_+ =10$ (right). The fundamental mode and first overtone are displayed as blue points.  }
\label{figcpII}
\end{center}
\end{figure}
%
% Figures
%

\begin{figure}[htbp]
\begin{center}
\epsfig{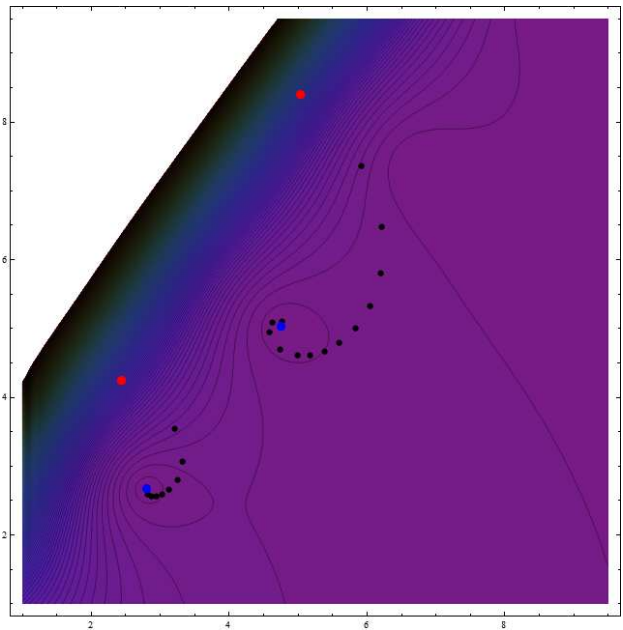}
\caption{Contour levels and evolution of $f(\omega)$ minima as function of $N$ for $r_+ =1$. The fundamental mode and first overtone are displayed as blue points. }
\label{figcpIII}
\end{center}
\end{figure}

In order to execute our numerical experiments, we implement the optimization method in \emph{Mathematica} without use any built-in function of the software. We employ the following parameters: $M = 250$ outer iterations, $m = 500$ inner iterations and $\epsilon = 10^{-5}$. All tests have been carried out on a single core of an Intel Core i5 CPU 2.5GHz with 4GB RAM running MAC OS X 10.9. To choose adequate values for lower and upper bounds $\ell, u$, we draw some contour levels of function $f(w)$ looking for the existence of global minimizers in an specific region of the complex $\omega$ plane.
As one can see in the Figures \ref{figcpII} and \ref{figcpIII} the minima of $f(\omega)$ are well distributed and linearly disposed. We named fundamental modes those QNM's with the lowest imaginary part and overtones all others.

In the Figure \ref{figcpIII}, the contour levels for $r_+=1$ are plotted. In this case the purple region is more extensive when compared with those the others black holes. This show us how difficult is to find the minimizers for small black holes. Since the neighborhood of the minimum is less located, becomes necessary to use a free-derivative numerical method.

In Table \ref{tab1} we list our numerical results and compare them with those obtained by Horowitz and Hubeny \cite{horowitz}. We have considered large and intermediate black holes varying the horizon radius from $r_+ = 100$ to $r_+ = 0.4$. The small black holes were omitted because even as Horowitz and Hubeny, we had problems with convergence of the QNM's values. The approximate values to the QNM's $\omega \in \C$ are indicated in the columns $\omega_R - i\ \omega_I$. The final values attained by the objective function of the problem (\ref{prob_otim}) are showed in the column $f(\omega^*)$. The columns $t(s)$ and $N$ indicate, respectively, the CPU time spent in each test and the quantity of terms considered in the truncated sum (\ref{fzn}). One can observe that our algorithm had results in good agreement with those obtained by Horowitz and Hubeny. Also, in our case, it is not necessary to consider a large number of terms in the truncated sum. The time of computation $t(s)$ indicates that our algorithm can be consider fast in to obtain solutions if $N$ is not so large.
%
% Table 1
%
\begin{table}[!htbp]
\begin{center}
\renewcommand{\arraystretch}{1.3}
\tabcolsep 10pt
{\scriptsize
\begin{tabular}{|c|r|rccc|}
\hline
\hline
\multicolumn{1}{|c}{} & \multicolumn{1}{|c}{Horowitz \& Hubeny} & \multicolumn{4}{|c|}{Luus-Jakola algorithm} \\
\hline
$r_+$ & \multicolumn{1}{c|}{$\omega_R - i ~\omega_I$} & \multicolumn{1}{c}{$\omega_R - i~\omega_I$} & $f(w^*)$ & $t(s)$ & $N$ \\
\hline
100 & $184.9534 - i~266.3856$ & $184.9530 - i~266.3860$ & 5.36E$-$07 & 31.88 & 50 \\
50 & $  92.4937 - i~133.1933$ & $ 92.4938 - i~133.1930$ & 1.98E$-$06 & 31.85 & 50 \\
10 & $18.6070 - i~26.6418$ & $18.6070 - i~26.6418$ & 8.09E$-$06 & 31.93 & 50 \\
5  &  $   9.4711 - i~13.3255$ & $9.4711 - i~13.3255$ & 2.89E$-$06 & 39.89 & 50 \\
1  &  $   2.7982 - i~2.6712$ & $2.7982 - i~2.6712$ & 5.48E$-$07 & 47.71 & 50 \\
0.8 & $  2.5878 - i~2.1304$ & $2.5877 - i~2.1304$ & 3.02E$-$06 & 39.71 & 50 \\
0.6 & $  2.4316 - i~1.5797$ & $2.4315 - i~1.5797$ & 2.29E$-$06 & 48.01 & 50 \\
0.4 & $  2.3629 - i~1.0064$ & $2.3629 - i~1.0064$ & 2.86E$-$06 & 140.46 & 90 \\
\hline
\hline
\end{tabular}}
\end{center}
\caption{The lowest QNM of a massless scalar field propagating in a Sch-AdS$_4$ for $\ell_* =0$ and several black holes sizes.}\label{tab1}
\end{table}

In addition, the QNM's for the massive scalar field propagating in \linebreak Schwarzschild-AdS$_4$ black hole was calculated using our derivative-free algorithm. Until we know these results are being presented by the first time here.

In Table \ref{tab2}, the QNM's for a massive scalar field for some values of mass are listed. We have considered large and intermediate black holes with horizon radius $r_+ = 100$, $r_+ = 10$ and $r_+=1$. The small black holes are omitted again by the same reasons presented above. For the three black holes analyzed both, the real and imaginary term of the fundamental mode and the first overtone grow when the mass increase. The presence of the mass term did not affect the stability of the black hole under scalar perturbations at least for the values of mass studied. Its influence in the relation between the black hole temperature $T_{H}$  and QNM's could not be established properly due to the low quantity of black holes analyzed. This issue will be addressed in a future work.
%
%
% Table 2
%
\begin{table}[htbp]
\begin{center}
\renewcommand{\arraystretch}{1.3}
\tabcolsep 10pt
{\scriptsize
\begin{tabular}{|c|c|c|}
\multicolumn{1}{c}{$r_+ =1$} & \multicolumn{1}{c}{Fundamental mode} & \multicolumn{1}{c}{First overtone} \\
\hline
$m$ & \multicolumn{1}{c|}{$\omega_R - i ~\omega_I$} & \multicolumn{1}{c|}{$\omega_R - i~\omega_I$} \\
\hline
0.0 & $2.7982 - i~2.6712$  & $4.7584 - i~5.0375$  \\
0.5 & $2.8721 - i~2.7690$  & $4.8359 - i~5.1334$  \\
1.0 & $3.0749 - i~3.0382$  & $5.0473 - i~5.3958$  \\
2.5 & $4.1096 - i~4.4183$  & $6.1090 - i~6.7293$  \\
5.0 & $6.3375 - i~7.3582$  & $8.3254 - i~9.5656$ \\
\hline
\hline
\noalign{\bigskip}
\multicolumn{1}{c}{$r_+ =10$} & \multicolumn{1}{c}{Fundamental mode} & \multicolumn{1}{c}{First overtone} \\
\hline
$m$ & \multicolumn{1}{c|}{$\omega_R - i ~\omega_I$} & \multicolumn{1}{c|}{$\omega_R - i~\omega_I$}  \\
\hline
0.0 & $18.6070 - i~26.6418$  & $31.8017 - i~49.1816$  \\
0.5 & $19.1146 - i~27.5782$  & $32.3219 - i~50.0993$  \\
1.0 & $20.5037 - i~30.1470$  & $33.7427 - i~52.6115$  \\
5.0 & $42.5554 - i~70.8348$  & $55.7932 - i~92.2299$  \\
10.0 & $75.7004 - i~129.8382$  & $88.2790 - i~150.5596$ \\
\hline
\hline
\noalign{\bigskip}
\multicolumn{1}{c}{$r_+ =100$} & \multicolumn{1}{c}{Fundamental mode} & \multicolumn{1}{c}{First overtone} \\
\hline
$m$ & \multicolumn{1}{c|}{$\omega_R - i ~\omega_I$} & \multicolumn{1}{c|}{$\omega_R - i~\omega_I$}  \\
\hline
0.0 & $184.9534 - i~266.3856$  & $316.1447 - i~491.6435$  \\
0.5 & $190.0025 - i~275.7443$  & $321.3163 - i~500.8150$  \\
1.0 & $203.8182 - i~301.4158$  & $335.4430 - i~525.9220$  \\
5.0 & $423.0855 - i~707.9776$  & $554.6813 - i~921.8352$  \\
10.0 & $752.5824 - i~1297.5387$  & $877.6441 - i~1504.6683$ \\
\hline
\hline
\end{tabular}}
\end{center}
\caption{The lowest QNM and the first overtone of a massive scalar field propagating in a Sch-AdS$_4$ for $\ell_* =0$ and three black hole sizes: $r_+ =1$ $r_+ =10$ and $r_+ =100$.}\label{tab2}
\end{table}

Finally, is important stress that these results for massive scalar field play an important role for the AdS/CFT once that its mass is intimately related to the conformal dimension of a quantum operator on the boundary.

%%%%%%%%%%%%%%%%%%%%%%%%%%%%%%%%%%%%%%%%%%%%%%%%%%%%%%%
\section{Conclusions and future work}\label{sec5}

In this work we have presented an interesting application of a derivative-free optimization method to compute QNM's of a massive scalar field evolving in intermediate and large Schwarzschild-AdS$_4$ black holes. We have implemented the algorithm in the software \emph{Mathematica} and compare the results obtained with the original results calculated by Horowitz and Hubeny. Our results showed to be in good agreement with those of the literature and it was not necessary to use a large number of terms in the truncated sum. The derivative-free algorithm proved to be an efficient alternative method to calculated QNM's in asymptotically AdS spacetimes.

In addition, the stability of the black hole under massless and massive scalar perturbations was confirmed. The application and discussion of these calculations for higher-dimensional AdS black holes is straightforward and it will appear in a future work. A possible extensions of this formalism applied for other matter fields evolving in asymptotically AdS black holes are under analysis.

%%%%%%%%%%%%%%%%%%%%%%%%%%%%%%%%%%%%%%%%%%%%%%%%%%%%%%%
\section*{Acknowledgements}

This work has been supported by CNPq (Conselho Nacional de Desenvolvimento Cient\' ifico e Tecnol\' ogico) and FAPEMIG (Funda\c{c}\~ao de Amparo \`a Pesquisa do Estado de Minas Gerais).





%%%%%%%%%%%%%%%%%%%%%%%%%%%%%%%%%%%%%%%%%%%%%%%%%%%%%%%

%
\end{document}